\documentstyle[psfig,prb,aps,twocolumn]{revtex}
\begin{document}
\draft

\twocolumn[\hsize\textwidth\columnwidth\hsize\csname %
@twocolumnfalse\endcsname

\title{Spinon signatures in the critical phase of the $\mathbf
(1,{1\over2})$ ferrimagnet in a magnetic field}

\author{A. K. Kolezhuk}
\address{Institute of Magnetism, National Academy of  
Sciences, 36(b) Vernadskii avenue, Kiev 252142, Ukraine\\
Institut f\"{u}r Theoretische Physik,
Universit\"{a}t Hannover, Appelstra{\ss}e 2, D-30167 Hannover, Germany}
\author{H.-J. Mikeska}
\address{Institut f\"{u}r Theoretische Physik,
Universit\"{a}t Hannover, Appelstra{\ss}e 2, D-30167 Hannover, Germany}
\author{K. Maisinger and U. Schollw\"ock}
\address{Sektion Physik, Ludwig-Maximilians-Universit\"at M\"unchen,
Theresienstr. 37, D-80333 Munich, Germany}

\date{December 18, 1998}

\maketitle

\begin{abstract}
We propose an effective theory for the critical phase of a quantum
ferrimagnetic chain with alternating spins $1$ and $1\over2$ in an
external magnetic field. With the help of the matrix product
variational approach, the system is mapped to a spin-$1\over2$ XXZ
chain in an (effective) magnetic field; as a byproduct, we obtain an
excellent description of the optical magnon branch in the gapped
phase.  Recent finite-temperature DMRG results for the low-temperature
part of the specific heat are well described by the present approach,
and the ``pop-up'' peaks, developing near the critical field values
and in the middle of the critical phase, are identified with the
contributions from two different spinon bands of the effective
spin-$1\over2$ chain. The effect should be as well observable in other
spin-gap systems in an external field, particularly in spin ladders.
\end{abstract}
]

\narrowtext


{\em Introduction.\/}
Recently, there has been a considerable interest in the properties of
one-dimensional (1D) quantum
ferrimagnets.\cite{Kahn,kmy97,bmy97+,AlcarazMalvezzi97,YamFuk98,Ivanov98,%
Yam+98,exp,Maisinger+98} A generic example of a 1D ferrimagnet is the
Heisenberg spin chain with antiferromagnetic nearest neighbor
interaction and alternating spins $1$ and $1\over2$, described by the
Hamiltonian 
\begin{equation}
\widehat H= J\sum_n (\mbox{\boldmath$S$\unboldmath}_n 
\mbox{\boldmath$\tau$\unboldmath}_n +
\mbox{\boldmath$\tau$\unboldmath}_n
\mbox{\boldmath$S$\unboldmath}_{n+1})
-h\sum_{n} (S_{n}^{z}+\tau_{n}^{z})\,,
\label{ham}
\end{equation}
where $\mbox{\boldmath$ S$\unboldmath}_n$ and
$\mbox{\boldmath$\tau$\unboldmath}_n$ are respectively spin-$1$ and
spin-$1\over2$ operators in the $n$-th elementary magnetic cell, and
$h=g\mu_{B}H$, where $H$ is the external magnetic field.
According to the Lieb-Mattis theorem, \cite{LiebMattis62} for $h=0$
the ground state of the system has total spin $S_{\rm tot}=L/2$,
where $L$ is the number of unit cells, and thus is necessarily
long-range ordered, making the problem amenable to the spin wave
theory (SWT) approach.\cite{bmy97+,Yam+98,Maisinger+98} 
Since the elementary cell consists of two spins, SWT yields two types
of magnons: a gapless ``acoustical,'' or ``ferromagnetic'' branch with
$S^{z}=L/2-1$, and a gapped ``optical,'' or
``antiferromagnetic'' branch with $S^{z}=L/2+1$. The optical
magnon gap was numerically found\cite{bmy97+} to be $\Delta_{\rm
opt}=1.759\,J$. The existence of two magnon branches manifests itself
in various thermodynamic quantities.\cite{YamFuk98,Yam+98} 

In magnetic field the acoustic branch acquires a gap,
while the optical gap decreases with the field.  If the field exceeds
the critical value $h=h_{c1}=\Delta_{\rm opt}$, the optical gap
closes, and the system enters the critical phase, which is expected to
be of the Luttinger-liquid type, in analogy with other models (e.g.,
spin-$1$ chain,\cite{affleck+90} gapped spin-${1\over2}$ chains and
ladders\cite{ChitraGiamarchi97,mila+}).  This phase extends up to the
second critical field $h=h_{c2}=3\,J$, where there is another
transition to the saturated ferromagnetic phase.\cite{Maisinger+98} In
the gapless regime $h_{c1}< h < h_{c2}$, the temperature dependence of
the specific heat has revealed a puzzling behavior \cite{Maisinger+98}
with a single well-pronounced low-$T$ peak which pops up around
$T\approx0.2\,J$ when $h$ is in the middle between $h_{c1}$
and $h_{c2}$; when $h$ is shifted towards $h_{c1}$ or $h_{c2}$, the
peak becomes flat and develops a shoulder with another ill-pronounced
peak centered at about $0.05\,J$.

In this paper, we show that in the critical phase the model
(\ref{ham}) can be mapped to an effective spin-$1\over2$ XXZ chain in
an external field, which yields quite an accurate description of
all the features of the low-temperature part of the specific heat.
The ``pop-up'' peak structure described above can be explained by
the presence of two different spinon bands of the effective
spin-$1\over2$ chain. This effect is rather general and should be
observable in other gapped 1D spin models, e.g., spin-$1\over2$ ladders.

{\em Effective model.\/} 
The general idea of any mapping to an
effective model is to reduce somehow the Hilbert space of the problem,
keeping only a few ``most important'' states per the elementary
cell. For example, in the strong-coupling limit of the spin-$1\over2$
ladder one keeps for each rung only the singlet and the lowest-energy
triplet component.\cite{mila+} For the $(1,{1\over2})$ ferrimagnet the
set of the cell wave functions $\psi_{jm}$ consists of a doublet
($j={1\over2}$) and a quartet ($j={3\over2}$). One would naively
expect that in a strong field it is now necessary to keep three states with
$(j,m)=({1\over2},{1\over2})$, $({3\over2},{1\over2})$,
$({3\over2},{3\over2})$; it is clear, however, that there is only {\em
one\/} ground state and {\em one\/} excitation becoming gapless at
$h=h_{c1}$, so that the interplay should be effectively between {\em
two\/} cell wavefunctions being ``proper'' linear combinations of
$\psi_{jm}$. Thus, the problem is how to identify those proper
combinations. The first step is to describe accurately the
optical magnons, since the critical phase is formed by their
condensation into the ground state.

The linear SWT captures the essential physics of the model at the
qualitative level;\cite{bmy97+} however, for a quantitative
description one has to go far beyond the linear approximation, keeping
the higher-order terms in $1/S$;\cite{Ivanov98,Yam+98} e.g., in linear
SWT $\Delta_{\rm opt}=J$, and the corrections from the leading $1/S$
term yield\cite{Yam+98} $1.676\,J$ instead of the numerical value
$1.759\,J$.  Thus we use a different scheme which has
proved\cite{kmy97} to be very successful for the $(1,{1\over2})$
ferrimagnet, namely the variational matrix
product\cite{Fannes+,Klumper+} (MP) approach.  In Ref.\
\onlinecite{kmy97} the following (non-normalized) variational ground
state wave function was proposed:
\begin{eqnarray} 
\label{gs} 
&& \Psi_{0} =\mbox{tr}\,(g_1 g_2 \cdots g_L),\\
&& g_{n}= u M^{0,{1\over2}} + v M^{1,{1\over2}}
+  M^{1,{3\over2}}\,, \quad
 M^{0,{1\over2}}=\openone \, \psi_{{1\over2},{1\over2}}\,,\nonumber\\
&& M^{1,{1\over2}}=
-{1\over\sqrt{3}}\sigma^{0} \psi_{{1\over2},{1\over2}}
+\sqrt{2\over3}\sigma^{+1}\psi_{{1\over2},-{1\over2}}\,,
\nonumber\\
&& M^{1,{3\over2}}=
{1\over\sqrt{2}}\sigma^{-1}\psi_{{3\over2},{3\over2}}
+{1\over\sqrt{6}}\sigma^{+1}\psi_{{3\over2},-{1\over2}}
-{1\over\sqrt{3}}\sigma^0\psi_{{3\over2},{1\over2}}\,.\nonumber
\end{eqnarray}
Here $\sigma^{0}=\sigma_{z}$ and
$\sigma^{\pm1}=\mp{1\over\sqrt{2}}(\sigma_{x}+i\sigma_{y})$ are the
Pauli matrices, $\openone$ is the 2$\times$2 unit matrix, $\psi_{jm}$
denote spin states of the $n$-th cell, and $u$, $v$ are the
variational parameters whose optimal values minimizing the energy of
(\ref{ham}) at $h=0$ were found\cite{kmy97} to be $u=1.3026$,
$v=1.0788$, with the variational ground state energy per unit cell
$E_{0}^{\rm var}=-1.449\,J$ (to compare with the numerical value
$E_{0}=-1.455$). It was shown that $\Psi_{0}$ possesses the correct
quantum numbers of $S_{\rm tot}= S^{z}_{\rm tot}=L/2$ and provides a
very good description of the ground state correlations.  Any quantum
averages for MP wave functions are readily calculated using the
transfer matrix technique; this nice formalism was first proposed in
Refs.\ \onlinecite{Fannes+,Klumper+} The MP approach is especially well
suited to this problem since the fluctuations are extremely
short-ranged, with the correlation radius smaller than one unit cell
length.\cite{kmy97,bmy97+}

\begin{figure}
\mbox{\psfig{figure=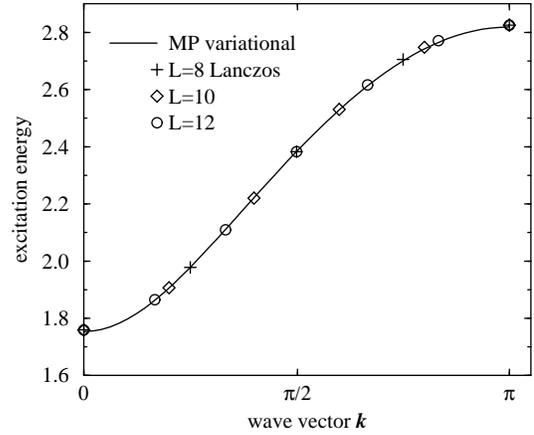,width=80mm,angle=-90}}
\vskip 5mm
\caption{\label{fig:afdisp}
Dispersion of the optical magnon branch (in $J$ units). The solid
line is variational result (\protect\ref{afdisp}), and symbols denote
the Lanczos data.  
}
\end{figure}

Now we construct the simplest trial wave function for the optical
magnon with the momentum $k$:
\begin{eqnarray} 
\label{opt} 
&& |k\rangle=\sum_{n}e^{ikn}|n\rangle,\quad
|n\rangle=\mbox{tr}\,(g_1 \cdots g_{n-1}\widetilde{g}_{n}g_{n+1}\cdots g_L),
\nonumber\\
&&  \widetilde{g}_{n}=\openone \, \psi_{{3\over2},{3\over2}} 
    + w\,\sigma^{+1}\, \psi_{{1\over2},{1\over2}}  + \sqrt{5/3}\,
        f\, Q^{1,{3\over2}},\\
&& Q^{1,{3\over2}}=\sqrt{3/5}\, \sigma^{0}\, \psi_{{3\over2},{3\over2}} 
 -\sqrt{2/5}\,\sigma^{+1}\,\psi_{{3\over2},{1\over2}}\,.\nonumber
\end{eqnarray}
The form of $\widetilde{g}_{n}$ in (\ref{opt}) is dictated by the
requirement that $|n\rangle$ is the state with $S_{\rm tot}=S^{z}_{\rm
tot}=L/2+1$, then, according to the general formalism presented in
Ref.\ \onlinecite{kmy97}, $\widetilde{g}_{n}$ should carry the
``hyperspin'' quantum numbers $({3\over2},{3\over2})$, while each 
 $g_{n}$ carries $({1\over2},{1\over2})$. Here $f$, $w$ are still
free (variational) parameters. Note that generally the states
$|n\rangle$ are not orthogonal to each other, but are orthogonal to
$\Psi_{0}$.  For the following, it is important to note that
$\widetilde{g}_{n}$ can be represented in the form
\[
\widetilde{g}_{n}= {f-1\over\sqrt{2}}\,g_{n}\, \sigma^{+1}
-{f+1\over\sqrt{2}}\, \sigma^{+1}\,g_{n}
+ \widetilde{w}\, \sigma^{+1} \,\psi_{{1\over2},{1\over2}} \, ,
\]
so that $|k\rangle$ in fact depends only on 
$\widetilde{w}=\big\{w+\sqrt{2}(u+{vf\over\sqrt{3}})\big\}$. 
Thus, one parameter in (\ref{opt}) is redundant, and we fix that
by choosing
\begin{eqnarray}
\label{f}
&& f=\big\{z+w\sqrt{6}(v-u\sqrt{3})\big\}/(2+z)\,,\\
&& z\equiv 2\sqrt{3}\,uv +\{ 3+12v^{2}(u^{2}+1)\}^{1/2}\,,\nonumber
\end{eqnarray}
which yields a remarkable property: the set of one-magnon 
states $\{|n\rangle\}$
becomes mutually {\em orthogonal,\/} considerably simplifying further
calculations. Moreover, one can show that arbitrary
$N$-magnon states $|n_{1},\ldots n_{N}\rangle$ become
orthogonal; however, they are not normalized, and, generally, $\langle
n,n+m|n,n+m\rangle$ does not coincide with $\langle n|n\rangle^{2}$, though
it tends very fast to the latter value as $m$ increases.
The one-magnon norm and the matrix elements of the Hamiltonian
(\ref{ham}) are given by
\begin{eqnarray} 
\label{mel}
&& {\cal N}_{0}
\equiv\langle n| n\rangle = 0.1173w^{2}-0.0337w+0.1432\,,\nonumber\\
&& \langle n| \widetilde{H} |n'\rangle
=Jx^{|n-n'|}(A+B\delta_{|n-n'|,1} +C\delta_{nn'})\,,\nonumber\\
&& x=0.2147,\quad A=-0.0483w^{2}+0.0808w-0.0284\,,\nonumber\\
&& B=-0.0839w^{2}+ 0.0379w +0.0010\,,\\
&& C=0.3359w^{2}-0.0495w+0.5860\,,\nonumber
\end{eqnarray}
where $\widetilde{H}\equiv \widehat{H}-L E_{0}^{\rm var}$. 
Now one can calculate the dispersion 
by minimizing the excitation energy $\varepsilon(k)=
\langle k| \widetilde{H}|k\rangle/ \langle k| k\rangle$,
\begin{equation} 
\label{afdisp} 
\varepsilon(k)= {J\over{\cal N}_{0}}\Big\{ C+2Bx\cos k 
+A{1-x^{2}\over 1+x^{2} -2x\cos k}\Big\}\,.
\end{equation}
In principle, the optimal value of $w$ should be calculated separately
for each $k$. One might first try to
minimize the gap $\Delta_{\rm opt}=\varepsilon(k=0)$, which yields 
$w=w_{0}=-3.8605$, with $\Delta_{\rm opt}^{\rm var}\simeq
1.754\,J$, the resulting dispersion  for $w(k)=w_{0}$ being in
excellent agreement with the exact diagonalization data
(see Fig.\ \ref{fig:afdisp}).\cite{comment}

Now, we can obtain the desired mapping by introducing effective
spin-$1\over2$ states $|\alpha_{n}\rangle$ at each cell,
$\alpha_{n}=\pm{1\over2}$, and making the identification
$|\alpha_{1}\alpha_{2}\cdots \alpha_{L}\rangle =
\mbox{tr}(R_{1}R_{2}\cdots R_{L})$, where $R_{n}=\widetilde{g}_{n}$,
$g_{n}$ for $\alpha_{n}={1\over2}$, $-{1\over2}$,
respectively. Introducing the spin-$1\over2$ operators
$s_{n}^{z,\pm}$, and restricting all effective interactions to nearest
neighbors only, one can write down the effective Hamiltonian for the
critical phase of the $(1,{1\over2})$ ferrimagnet (\ref{ham}):
\begin{eqnarray} 
\label{Heff} 
&& \widehat{H}_{\rm e}=\sum_{n} {J_{xy}\over2}(s_{n}^{+}s_{n+1}^{-} 
+s_{n}^{-}s_{n+1}^{+}) +J_{z} s_{n}^{z} s_{n+1}^{z} -2h_{e}
s_{n}^{z}, \nonumber\\
&& J_{xy}=-2t_{1},\quad   J_{z}=U_{1},\quad 
h_{e}=(h-t_{0}-U_{1})/2 ,\\
&& t_{m}\equiv {\langle n|\widetilde{H}|n+m\rangle \over {\cal N}_{0}},\;
U_{m}\equiv {\langle n,n+m|\widetilde{H}|n,n+m\rangle \over 
\langle n,n+m|n,n+m \rangle}-2 t_{0}.\nonumber
\end{eqnarray}
Here a remark is in order. 
From (\ref{mel}) one can see that the hopping amplitudes $t_{m}\propto
x^{m}$ are very small for $m\geq 2$, $t_{2}=-0.0242$, $t_{3}=-0.0052$,
$\ldots$, which justifies keeping the terms of the
$s_{n}^{+}s_{n+m}^{-}$ type only up to $m=1$ in (\ref{Heff}). The same
reasoning holds for neglecting  other  interaction terms
like $(s^{+}_{n}s^{+}_{n+1}s^{-}_{n-1}s^{-}_{n+2})$,
$(s^{+}_{n}s^{+}_{n+2}s^{-}_{n}s^{-}_{n+2})$, etc.  

The numerical values for the parameters $t_{0}$, $t_{1}$, $U_{1}$,
which determine our effective model, are $t_{0}=2.3370$,
$t_{1}=-0.2608$, $U_{1}=0.1187$, so that the spin-$1\over2$ chain
(\ref{Heff}) is in a gapless XY phase with $\Delta=J_{z}/J_{xy}\simeq
0.227$. At the critical value\cite{Korepin+} of the field
$h_{e}=h_{e,c}={1\over2} J_{xy}(1+\Delta)$ the spin-$1\over2$ chain undergoes
a transition into the saturated ferromagnetic phase. The obvious
symmetry $h_{e}\mapsto -h_{e}$ corresponds for the ferrimagnet to the
symmetry around $h_{0}={1\over2}(h_{c1}+h_{c2})$. Such a symmetry was indeed
observed in Ref.\ \onlinecite{Maisinger+98} for the behavior of the
specific heat, though for the $h<h_{0}$ side it was somewhat plagued by
the low-$T$ contribution of the acoustical magnon branch. 
In what follows, we compare our results only with the data for
$h>h_{0}$.  Because of cutting out non-nearest neighbor interactions,
the critical fields for the model (\ref{Heff}) are slightly different
from the real ones: $h_{e}=\pm h_{e,c}$ corresponds to
$h=h_{c1}'\simeq1.81$ and $h=h_{c2}'\simeq3.09$ instead of the correct
values listed above.  Thus, for a comparison to the numerical data
below, we have linearly rescaled the field interval
$(h_{c1}',h_{c2}')$ onto $(h_{c1},h_{c2})$.

\begin{figure}
\mbox{\psfig{figure=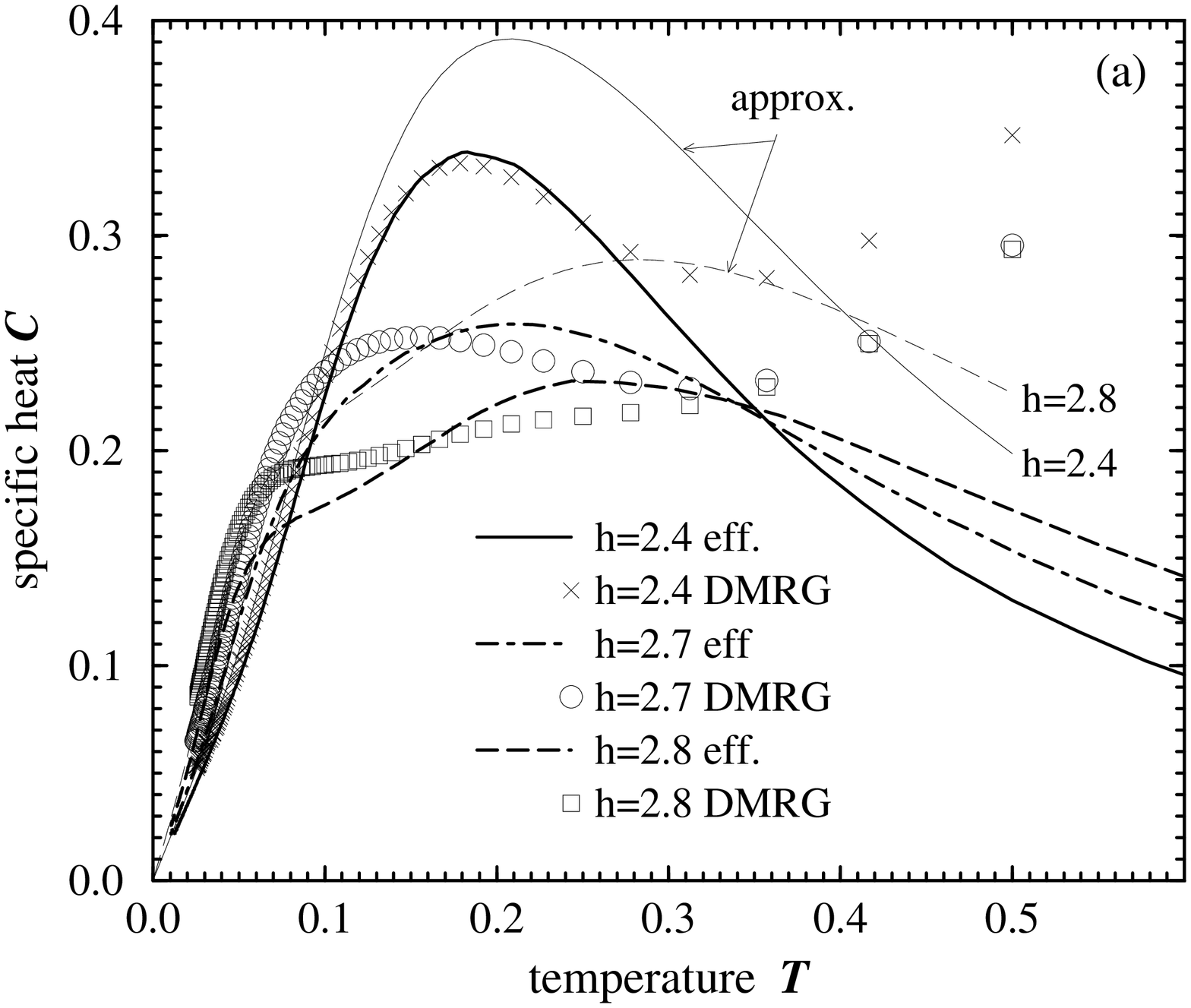,width=82mm,angle=-90}}
\vskip 0.1pt
\mbox{\psfig{figure=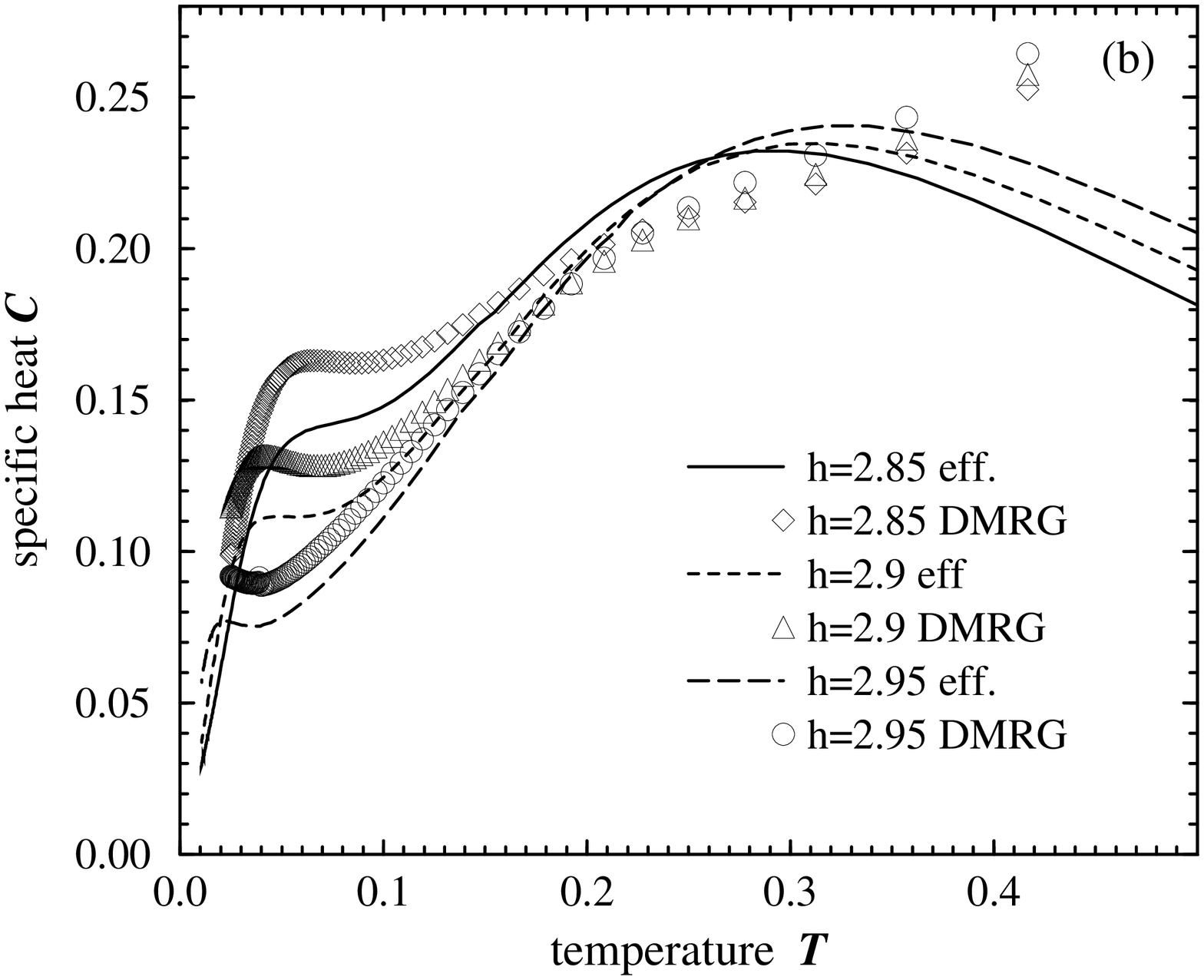,width=82mm,angle=-90}}
\vskip 2mm
\caption{\label{fig:specheat} Temperature dependence of the specific
heat for the model (\protect\ref{ham}).
Symbols denote the DMRG data, and thick curves
correspond to the Bethe ansatz result (\protect\ref{heat}) for the
effective model (\protect\ref{Heff}). Two thin curves in (a) illustrate
the first-order perturbative result based on
(\protect\ref{jw}).
}
\end{figure}

{\em The specific heat.\/}
The temperature dependence of the specific heat for
the spin-$1\over2$ chain can be calculated using Kl\"umper's
version\cite{Klumper-TB} of the thermodynamic Bethe ansatz; this
amounts to solving numerically the following system of two coupled
nonlinear integral equations:
\begin{eqnarray} 
\label{klump}
y_{1}&=&{\pi\beta\over \gamma\cosh{\pi x\over\gamma}} -{\pi \beta
\widetilde{h}\over 2 (\pi-\gamma)}
-\int_{-\infty}^{+\infty}\!\! dx' \Big\{
K(x-x')L(y_{1})\nonumber\\ 
 &-&K(x-x'-i\gamma+i\delta)L(y_{2})\Big\}\, ,
\end{eqnarray}
the other equation being of the same form with
$i\to -i$, $y_{1}\leftrightarrow y_{2}$, and $\widetilde{h}\to -
\widetilde{h}$.  Here $L(y)=\ln(1+e^{-y})$, $\Delta=\cos\gamma$,
$\delta$ is an infinitesimal positive number,
$\widetilde{h}=4h_{e}/(J_{xy}\sin\gamma)$, $\beta=2T/(J_{xy}\sin\gamma)$, and
the kernel $K(x)$ is defined as follows:
\begin{equation}
\label{kernel}
K(x)={1\over4\pi}\int_{-\infty}^{+\infty} dq {\sinh[(\pi/2-\gamma)q]\,
\cos qx\over \cosh (\gamma q/2)\sinh[(\pi-\gamma)q/2]}\,.
\end{equation}
(Note that the last term in (\ref{klump}) contains a singularity!).
 Then the specific heat $C(T)$ can be found as
\begin{equation}
\label{heat}
C=\beta^{2}{\partial^{2} f\over\partial \beta^{2}},\quad 
f={1\over2\gamma}\int_{-\infty}^{+\infty}dx\,{L(y_{1})+L(y_{2})\over
\cosh(\pi x/\gamma)}
\end{equation}
In Fig.\ \ref{fig:specheat} we compare the results given by (\ref{heat})
with the numerical data for the low-temperature part
of $C(T)$ for the ferrimagnet (\ref{ham}), obtained by means of the
recently proposed\cite{trm-DMRG} transfer-matrix density matrix
renormalization group (DMRG) technique. The agreement is generally
satisfactory, and becomes very good near $h=h_{0}$. The ``pop-up''
peak behavior is clearly reproduced.

Qualitatively, the behavior of $C(T)$ can be understood without
appealing to the machinery of the Bethe ansatz. Passing to the
Jordan-Wigner fermions in (\ref{Heff}), and applying the first-order
perturbation theory in $\Delta$, one obtains the following simple
expression for the fermion dispersion $\epsilon(k)$:
\begin{eqnarray} 
\label{jw}    
 \epsilon(k)&=&\Delta-\lambda-\cos k \\
&-&(2\Delta/\pi)\,\theta(1-\lambda)\big\{\arccos\lambda
+(1-\lambda^{2})^{1/2} \cos k\big\}\,,\nonumber 
\end{eqnarray} 
here $\epsilon(k)$ is measured in $J_{xy}$ units,
$\lambda=2h_{e}/J_{xy}$, and $\theta$ is the Heaviside function. The
 specific heat can be trivially calculated; in Fig.\
\ref{fig:specheat}a, we have included two curves obtained
within the above approximation. One can see that, despite the general
tendency to overestimate $C$, this simple approach nevertheless
captures the essential physics. In zero field the
contributions into the specific heat from particles and holes are
equal;  with increasing field, the hole bandwidth grows up, while the particle
bandwidth decreases, and the average band energies do not
coincide. This leads to the presence of two peaks in $C(T)$: holes
yield a strong, round peak moving towards higher temperatures with
increasing the field, and the other peak (due to the particles) is
weak, sharp, and moves to zero when $h_{e}$ tends to $h_{e,c}$. In the
spinon language, one has essentially the same picture for the two bands of
spinons with opposite spin (the fact first noticed by
Kl\"umper\cite{Klumper-TB}).


{\em Summary.\/} An effective theory for the critical phase of a
quantum ferrimagnet with alternating spins $1$ and $1\over2$ in
external magnetic field is proposed.  Using the matrix-product
formalism, we have mapped the low-energy sector of the system to the
spin-$1\over2$ XXZ model in  an effective magnetic field; as a
byproduct, we obtain an excellent description of the optical magnon
branch in the gapped phase.  Recent transfer matrix DMRG
results\cite{Maisinger+98} for the specific heat $C(T)$ of the
ferrimagnet find natural explanation within the present approach. The
origin of ``pop-up'' peaks observed near the critical field values and
in the middle of the critical phase, is clarified: they can be
identified with the contributions from two different spinon bands of
the effective spin-$1\over2$ chain.  We believe those results should
be accessible to experimental verification, and we would like to
emphasize that this effect is {\em general} and should manifest itself
in the critical regime of essentially {\em any\/} gapped 1D spin
system in a magnetic field, as far as it is possible to construct the
mapping to a spin-$1\over2$ chain. Spin-$1\over2$ ladder compounds like
$\rm Cu_{2}(C_{5}H_{12}N_{2})_{2}Cl_{4}$ would be the natural
candidates.\cite{mila+,chaboussant+98}

{\em Acknowledgements.\/} This work was supported by the German
Federal Ministry for Research and Technology (BMBFT) under the
contract 03MI5HAN5.  A.K. gratefully acknowledges the hospitality
of the Hannover Institute for Theoretical Physics and support from the
Ukrainian Ministry of Science under the Grant No. 2.4/27.

\newpage

\ 

\vskip -18mm

\vskip -5mm


\begin{references}
\bibitem{comment} No finite-size extrapolation was
necessary due to a very short correlation length $\xi\simeq
0.365$.\protect\cite{kmy97}
\end{references}
\end{document}